# Emissivity of freestanding membranes with thin metal coatings

P. J. van Zwol[1*], D.F. Vles[1], W.P. Voorthuijzen[1], M. Péter[1], H. Vermeulen[1], W.J. van der Zande[1]
J. M. Sturm[2], R.W.E. van de Kruijs[2], F. Bijkerk [2]

[1] ASML Netherlands B.V., De Run 6501, 5504 DR Veldhoven, The Netherlands
[2] Industrial Focus Group XUV Optics, MESA+ Institute for Nanotechnology, University of Twente, PO Box 217, 7500 AE, The Netherlands

*Pieter-jan.van.zwol@asml.com

Freestanding silicon nitride membranes with thicknesses down to a few tens of nanometers find use as TEM windows or soft X-ray spectral purity filters. As the thickness of a membrane decreases, emissivity vanishes, which limits radiative heat emission and resistance to heat loads. We show that thin metal layers with thicknesses in the order of 1 nm enhance the emissivity of thin membranes by two to three orders of magnitude close to the theoretical limit of 0.5. This considerably increases thermal load capacity of membranes in vacuum environments. Our experimental results are in line with classical theory in which we adapt thickness dependent scattering terms in the Drude and Lorentz oscillators.

## Introduction

Thin membranes are routinely used as (soft) X-ray spectral purity windows [1]. Specifically in the EUV regime their thickness is limited to less than 100 nm if a high transmission is desired [2]. Due to a compatibility with silicon related deposition and etch processes, SiN or Si membranes are often used as windows for transmission electron microscopy and X-ray diffraction elements. X-ray or electron absorption in these windows may lead to heating due to the low heat capacity and in plane conduction of these membranes. An issue for semiconductor based membranes and pellicles such as SiN and Si, is the vanishing emissivity with decreasing membrane thickness (Fig. 1). Specifically under vacuum conditions the low emissivity limits the ability of these membranes to release heat. Figure 1 shows how at room temperature, the emissivity of the dielectrics decreases when the thickness decreases. Metals behave differently, for example, gold (Au) and ruthenium (Ru) reveal a sudden increase in emissivity below roughly 100 nm.

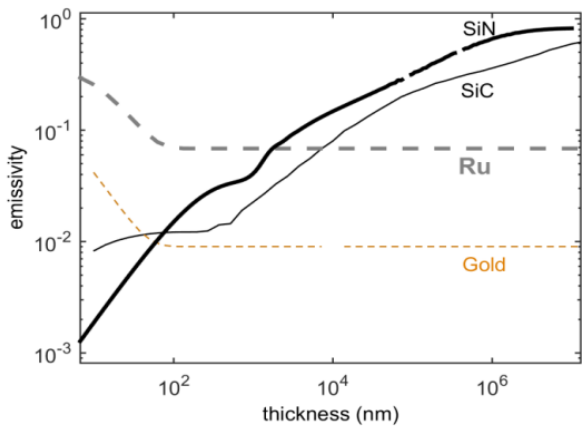

Figure (1): Room temperature emissivity versus thickness for a few metals and dielectrics. The values have been calculated from the optical constants of the materials of Palik [11] and methods in ref. [10].

As noticed in [3] for thin Pt films, the absorption or emissivity of metals increases if they are thin with respect to the infrared skin depth. The absorption coefficient of metals is relatively high and an etalon effect enhances the field inside the metal film, which also reduces the reflectivity of the vacuum- metal interface. The near room temperature emissivity increases when the thickness becomes comparable to the infrared skin depth of ~10 nm [4]. The maximum theoretical absorbance of a metal is close to 0.5 in the infrared region (1-100µm) in the situation when the refractive index (n) and extinction coefficient (k) have similar values [5], which is the case for metals in the infrared. Thin metal layers with enhanced absorption are applied in infrared sensors [6]. The goal of the present study is to understand what the optimum thickness of the metal film is on freestanding SiN membranes for achieving maximum emissivity and improved heat resistance.

## Sample preparation

Ruthenium, (Ru) coatings were deposited on freestanding square 5x5 mm$^2$ and 10x10 mm$^2$ silicon nitride membranes of only 25nm thickness. Ru was chosen for two reasons. Ru is relatively resistant to oxidation and Ru has limited inter-diffusion in SiN. Ru is known to form closed films around 1nm thickness [7]. Gold (Au) for example does not easily grow closed films and a percolation threshold is observed at 5 nm [8]. Thin 25nm SiN membranes are almost completely transparent in the visible to infrared range, apart from an absorption peak around 12µm. The combination of SiN membranes and Ru films gives a unique opportunity to measure the transmission and reflection and deduce the absorption of metallic films as it changes with thicknesses far below the skin depth. Such information cannot be obtained when using thick substrates because in this case only reflectivity can be measured.

In the experiments presented in this paper, Ru was coated on the SiN membranes by DC magnetron sputtering in an ultra-high vacuum deposition system. The chemical composition was analyzed with X-ray Photoelectron Spectroscopy (XPS), employing a ThetaProbe instrument from Thermo Scientific. Figure 2 shows XPS spectra (for a take-off angle of 34.25° with respect to the sample normal) and fitted components of elemental Ru and Ru oxide for selected layer thicknesses, as well as fitted amounts of Ru metal and oxide from the Ru 3d spectral region (inset) for all deposited thicknesses. The results show that ~0.5 nm of deposited Ru is consumed for forming Ru oxide (probably formed after exposure to atmosphere). For larger Ru thickness oxidation of the ruthenium does not measurably increase. XPS unfortunately does not reveal silicide formation due to the absence of measurable shifts (see below).

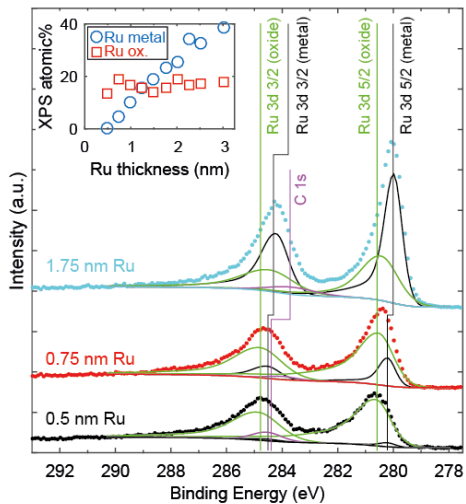

Figure (2): Ru 3d XPS spectra of membrane samples coated with indicated Ru thickness. Measured data is represented with dots, fitted components for Ru metal, Ru oxide and carbon with black, green and purple lines, respectively. The inset shows quantified amounts of metallic and oxidized Ru for all deposited layer thicknesses.

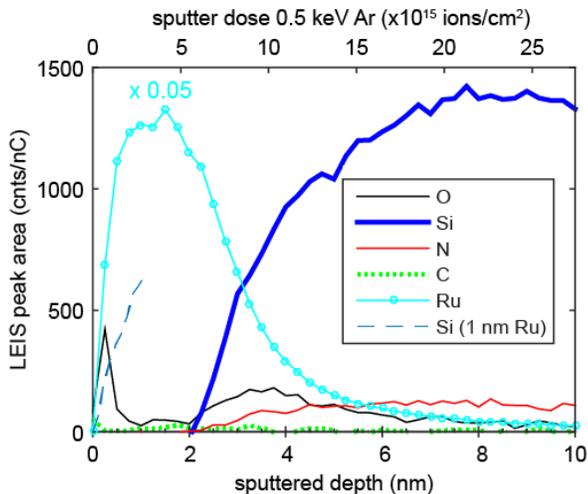

Figure (3): LEIS sputter depth profile of a membrane sample coated with 3nm Ru (solid lines). The sputtered depth is indicated based on the calibrated sputter rate for ruthenium. The dashed line show the initial Si signal from a sputter depth profile of a sample coated with 1 nm Ru, as confirmation of the calibration of the sputter depth scale.

In order to find out the thickness required for forming a closed Ru layer, selected samples were analyzed with Low-Energy Ion Scattering (LEIS), using a Qtac-100 instrument from ION-TOF. Figure 3 shows a sputter depth profile of a sample coated with 3 nm Ru (solid lines), which was obtained by alternately measuring the surface composition with a 3 keV He$^+$ beam (1×1 mm$^2$ area, normal incidence) and sputtering a 2×2 mm$^2$ area with 0.5 keV Ar$^+$ ions at 59° angle of incidence with respect to the surface normal. On the vertical axes, the peak area of the LEIS surface peaks is plotted, which indicates the composition of the outermost atomic layer [9]. The data shows that after removal of 2nm Ru by sputtering, a Si surface peak appears. From this, it is suggested that at least 1nm of Ru is needed for forming a closed layer on the SiN membrane samples. For confirmation, the sputter depth profile was also measured for a 1nm Ru coated sample. For this sample the Si signal is plotted with a dashed line in figure 3. In this case the Si signal (initially being zero) rises immediately. For the 1nm Ru film, the rise of the Si signal at 0nm sputter depth is similar to the 3nm Ru film at a sputter depth of 2nm. This confirms that the sputter depth scale was correctly calibrated and that 1nm Ru is indeed the onset for forming a closed layer. It should be noted that this is a conservative estimate, because the sputter depth profiling may induce some intermixing, thereby displacing Si from deeper layers to the surface.

The finding that 1 nm of Ru has to be deposited for forming a closed layer on top of the membrane, can indicate that Ru grows in 3D islands and/or that Ru forms a silicide-like intermixed layer with the membrane. It should be noted that LEIS measurements cannot distinguish between these two cases. Unfortunately, in XPS no significant binding energy shift of the Ru 3d peak upon formation of a silicide was found. Also for deposition of Ru on amorphous Si, where silicide formation is known to occur, no shifts in the Ru peaks were measured (unpublished work). Atomic Force Microscopy (AFM) showed no measurable increase of the surface roughness of the Ru covered membranes relative to the non-coated membranes, which showed a roughness of 0.31 ± 0.05nm root-mean-square (RMS) for a 1×1μm$^2$ image. This suggests that the threshold for forming a closed Ru layer relates to intermixing of Ru and SiN.

**Optical properties and emissivity**
Absorbance (A) can be determined from the reflectance (R) and transmittance (T) and the energy balance as A=1-R-T. From Kirchoff's law the emissivity must equal absorbance ε=A at any angle and wavelength. All these variables depend on the wavelength ω. We have measured R and T in the range 1-15μm with an FTIR under normal incidence and in the range 0.2-1.7μm with a Woollam ellipsometer. In the ellipsometer, T was measured under normal incidence and R under 20 degrees from normal incidence.

Measurements have been performed for 13 thicknesses of Ru between 0 and 3 nm over the full wavelength range as well as thicknesses of 4 and 8 nm FTIR data for the longer wavelengths. Figure 4 shows a representative set of measured data as well the as the predicted behavior based on optimization of a limited number of parameters describing the optical parameters of thin film Ru. For reference also Palik's data for bulk Ru is shown for optically thick 50nm Ru on SiN.

This difference in measurement angle in the ellipsometer gives a small systematic error in the derived absorption as can be seen from figure 4c.

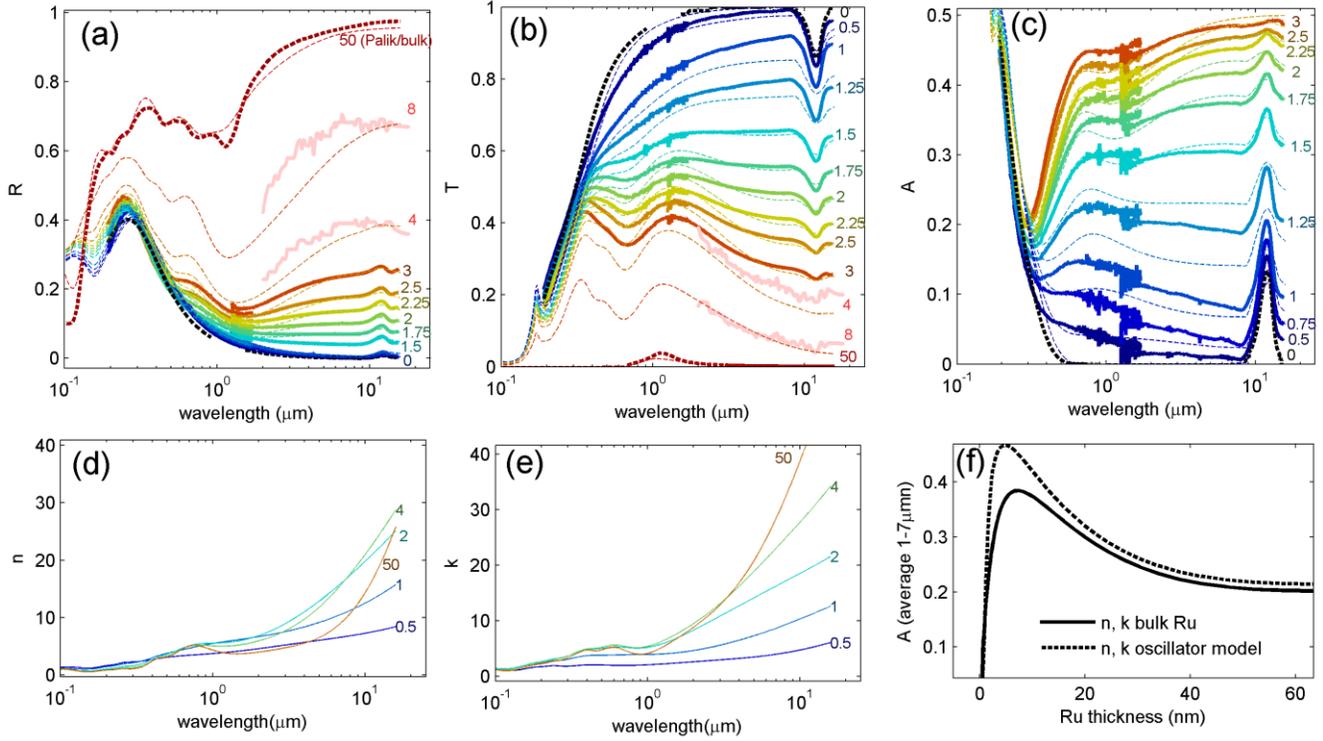

Figure (4): (a-c) FTIR and Ellipsometer data for reflectance, transmittance and absorbance (thick lines). The 50nm data is from Palik data for bulk Ru [11]. The thin dashed lines are the fitted Lorentz and Drude oscillator model from the text. Ru thicknesses are indicated in nm. (d,e) Derived optical constants (See model of table 1) for Ru films, as obtained from experimental data for different indicated thicknesses in nm and the wavelength range 100nm-15μm. (f) Average absorbance in 2-7 micron range calculated using bulk Ru optical constants (solid) and with the optical constants from (d) and (e) (dashed).

Around 1μm the FTIR and ellipsometer results do not exactly overlap. Nearly all infrared properties refer to the Ru thin film properties as SiN has no measurable absorption in this region apart from the absorption peak found around 12μm. We observe that only 0.5 nm of deposited Ru enhances the absorption over the entire (near) IR range significantly close to 0.1. A maximum absorption of 0.48 is reached for 3 nm thick Ru films, close to the theoretical maximum for freestanding metal films. For wavelengths above 10 microns both the measured absorption and transmission are close to 0.25, and n and k of the Ru film are nearly equal, in agreement with theoretical findings in ref. [5].

We model the SiN+Ru stack as a multilayer using Fresnel reflection and transmission coefficients [10]. In the visible and infrared region the complex dielectric function of metal films can be approximated by a set of Lorentz oscillators for inter- and intra-band transitions and a Drude model for the conduction electrons. We write the dielectric function as $\varepsilon(\omega)=\varepsilon(\omega>\omega_{duv})+\varepsilon_{Lorentz(1)}+...+\varepsilon_{Lorentz(n)}+\varepsilon_{Drude}$. A Lorentz oscillator is defined as:

$$\varepsilon_{Lorentz} = \frac{\omega_p^2}{\omega_0^2 - \omega^2 - i\omega_\tau \omega}$$

Here $\omega_0$ is the oscillator frequency, $\omega_p$ is the free electron plasma frequency and $\omega_\tau$ is the electron scattering frequency. The Drude model is a Lorentz term with $\omega_0=0$. We find that the complex dielectric function in this paper for Ru can be written in the form $1+\varepsilon_{Lorentz}(1)+....+\varepsilon_{Lorentz}(n)+\varepsilon_{Drude}$.

The bulk dielectric function of Palik [11] (used for the 50 nm Ru film in figure 4) doesn't describe the optical response of Ru with thicknesses below 20 nm. Ru has typically a 0.5nm $RuO_x$ layer ambient conditions. We model the Ru + $RuO_x$ system on SiN as a two layer system without roughness or inter-diffusion layers. Here the Ru+$RuO_x$ layer is treated as a single 'effective' layer. The optical properties of our SiN film were obtained experimentally using uncoated SiN membranes. Instead of fitting individual measurement data for each Ru thickness we fit a single thickness dependent dielectric function using a set of 6 Lorentz oscillators and a Drude term; 5 oscillators for the VIS-DUV range 1 for the NIR and 1 Drude term for the FIR. Oscillator values are shown in table 1.

The resulting fits and the thickness dependent complex refractive index n and k are shown in figure 4. We stress that the oscillator function converges to the measured bulk data of Palik for thick Ru films. For thinner Ru films our fits reveal an expected increasing electron scattering frequency when the Ru film thickness becomes comparable to the electron mean free path. For the Lorentz oscillator in the mid infrared with $\omega_0=0.5$eV (around 2 micron) we empirically find also an increase in strength ($\omega_p$) when fitting the data for thicknesses around 1.5-4 nm. A similar effect was observed for gold films [8] where the Drude model was not sufficient to model the response of thin gold layers around 1-2µm.  Figure 4a-c shows that the optimized set of oscillator parameters describe the reflection and transmission reasonably well. For the thinnest Ru films <1.5nm the deviations are slightly larger. This can in principle be made to fit better by adding more and higher polynomial terms of $\omega_\tau$ versus thickness, but we believe not much is learned from this. Our fit mainly serves to show that classical theory is enough to explain the result, and the oscillator function gives easily transferable optical constants for use by others.

Note also that the Fermi wavelength of Ru is about 0.5 nm.  In ideal situations, extra oscillations of R and T versus thickness d on the order of the Fermi wavelength may be expected related to quantum confinement. These oscillations are not seen in our data [12] due to the 0.5 nm of RuOx and the 0.3nm RMS roughness present, both of comparable size to the Fermi wavelength.

Figure 4 also shows the resulting optical constants for our thin Ru layer samples. They are clearly different from the bulk values. The results also show that the real and imaginary parts are nearly equal, hence promoting a maximum emissivity. The optical constants do not change monotonously with the thickness, at some wavelengths one finds a maximum at certain thickness. Figure 4(f) shows that the increased electron scattering has a distinct positive effect of enhancing the emissivity of thin Ru films.

| Oscillator | $\omega_0$ (eV) | $\omega_p$ (eV) | $\omega_\tau$ (eV) |
|---|---|---|---|
| Lorentz 1 | 13 | 13 | 6 |
| Lorentz 2 | 6.5 | 4.7 | 2.2 |
| Lorentz 3 | 5 | 4.7 | 1.7 |
| Lorentz 4 | 3 | 7.1 | $1.1+1/d$ |
| Lorentz 5 | 1.95 | 9.2 | $1.15+0.7/d+1/d^3$ |
| Lorentz 6 | 0.5 | $(30+180/d^{0.7})^{0.5}$ | $1.2+2/d+3/d^2+5/d^3$ |
| Drude | 0 | 5.8 | $0.088+0.4/d+2/d^3$ |

Table (1): The fitted oscillators that describe the dielectric function (see also text). This set of oscillators is valid in the range 0.1-15µm and is a function of Ru thickness d (nm) for metallic Ru grown on SiN in the thick Ru limit it converges to Paliks data for bulk Ru.

### Heat load tests
We also subjected the membranes to a heat load in order to confirm the increase in emissivity and to demonstrate the ability of the coated membranes to sustain larger heat loads. The membranes were exposed in vacuum to extreme UV light from an Energetiq Xenon EUV source. The base pressure in the exposure chamber is $10^{-8}$ mbar, but increases to $10^{-4}$ mbar due to Xenon gas present when the source is on. Under these low pressure conditions, the equilibrium temperature of the membrane is determined by the absorbed EUV light and by radiative cooling. The light of the EUV source was focused unfiltered on the membrane. Exposures of EUV-sensitive foils with sensitivity in the 10-20 nm range indicated a maximum power density of 5 W/cm$^2$ in a spot with a full width at half maximum (FWHM) of 1.5 mm. We estimate using Xenon spectra measured by Kieft et al. [13] and EUV optical data from the CXRO data base [14], that the membranes absorbed about 1.5 W/cm$^2$.

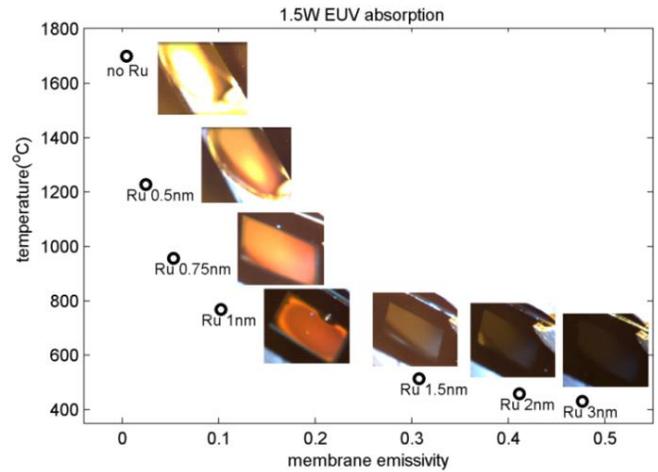

Figure (5): Calculated membrane temperature under EUV load using measured optical data. About 1.5W/cm$^2$ EUV is absorbed in the center of the spot.  Photos of the membranes under exposure conditions are also shown. The Xenon EUV source also outputs some white-blue visible light that can be seen reflected at the edges.

Uncoated SiN membranes were glowing white hot and broke in a few seconds. White hot glow indicates temperatures close to or in excess of 2000$^o$K which is close to the melting temperature of SiN. The coated membranes showed a red hot glow for Ru thicknesses up to 1 nm. For Ru thicknesses above this range no glow was observed. Figure 5 shows that the color of the emitted light from the membranes correlates well with the expected equilibrium temperature. This temperature was calculated from the balance between absorbed EUV power and Stefan-Boltzmann law for IR emission, were the emissivity was obtained from Planck's law and the measured absorption spectrum.

### Conclusion
By depositing very thin metal layers on thin SiN membranes, we have been able to determine the optical properties directly to assess the emissivity. We have shown that thin metal layers can drastically reduce the temperature of a thin membrane under heat load from above 1500$^o$C to more manageable temperatures of 400$^o$C. We found that increased electron scattering enhanced emissivity more than expected from theory using bulk optical constants. We believe the effect is generally true for all metals or semimetals with Drude conductivity. Future work could include investigation of quantum confinement of the electron gas if interlayer roughness can be reduced to angstrom levels. Conductive ceramics with similar thermo-mechanical properties as SiN could be interesting as a replacement for SiN. Finally we think that the effect of ultra-thin metals on Casimir Forces and Near Field Heat transfer is worth investigating. Besides ultra-thin metals on SiN membranes, thin metal coatings on other substrates such as KBr and KCl that are optically transparent in the infrared may have some interest.

**Acknowledgement**

The work of the XUV group is financially supported by ASML, Carl Zeiss SMT and the Stichting voor Fundamenteel Onderzoek der Materie (FOM).

[1] S.Labov, S.Bowyer, and G.Steele, Applied optics 24, 576 (1985) doi: 10.1364/AO.24.000576

[2] L. Scaccabarozzi ; D. Smith ; P.R. Diago ; E. Casimiri ; N. Dziomkina ; H. Meijer; Proc. SPIE Extreme Ultraviolet (EUV) Lithography IV, 867904 (2013) doi:10.1117/12.2015833

[3] G. D. Mahan and D. T. F. Marple, Appl. Phys. Lett. 42, 219 (1983) doi: 10.1063/1.93898

[4] S. Edalatpour and M. Francoeur, *Journal of Quantitative Spectroscopy and Radiative Transfer* **118**, 75 (2013) doi:10.1016/j.jqsrt.2012.12.012

[5] L. N. Hadley and D. M. Dennison, J. Opt. Soc. Am. 37, 451 (1947) doi: 10.1364/JOSA.37.000451

[6] S. Bauer, S. Bauer-Gogonea, W. Becker, R. Fettig, B. Ploss, W. Ruppel, W. von Munch, Sensors and Actuators A 37, 497 (1993) doi:10.1016/0924-4247(93)80085-U

[7] S. Bajt, et al, Applied optics 42, 5750 (2003) doi: 10.1364/AO.42.005750

[8] T.Brandt, M. Hövel, B. Gompf, and M. Dressel, Phys. Rev. B 78, 205409 (2008) doi:10.1103/PhysRevB.78.205409

[9] H. H. Brongersma, M. Draxler, M. de Ridder and P. Bauer, Surf. Sci. Rep. 62 (3), 63-109 (2007) doi:10.1016/j.surfrep.2006.12.002

[10] D. L. Windt, Comput. Phys. 12, 360 (1998) doi:10.1063/1.168689

[11] D.E. Palik, *Handbook of optical constants of solids*. Vol. 3. Academic press (1998)

[12] L.A. Kuzik, V.A. Yakovlev , F.A. Pudonin, G. Mattei, Surface Science 361 882 (1996) doi:10.1016/0039-6028(96)00556-0

[13] E. R. Kieft, K. Garloff, J. J. A. M. van der Mullen, and V. Banine, Phys. Rev. E 71, 036402 (2005) doi:10.1103/PhysRevE.71.036402

[14] B.L. Henke, E.M. Gullikson, and J.C. Davis. Atomic Data and Nuclear Data Tables 54 181 (1993) doi:10.1006/adnd.1993.1